\documentclass[twocolumn,aps,prb]{revtex4-1}

\usepackage{times}
\usepackage{graphicx}
\usepackage{amsfonts}
\usepackage{amsmath, amsthm, amssymb}
\usepackage{dsfont}
\usepackage{mathptm}
\usepackage{color}
\usepackage{subcaption}
\usepackage{standalone}
\usepackage{pgfplots}
\pgfplotsset{compat=1.13}
\usepackage{tikz}
\usetikzlibrary{decorations.markings}
\usetikzlibrary{shapes,positioning,arrows}
\usepackage{bm}
\usepackage[colorlinks=true, allcolors=black, citecolor=blue, linkcolor=blue]{hyperref}

\renewcommand{\i}{\mathrm{i}}
\newcommand{\cre}[1]{c^{\dagger}_{#1}}
\newcommand{\anni}[1]{c_{#1}}
\newcommand{\acre}[1]{a^{\dagger}_{#1}}
\newcommand{\aanni}[1]{a_{#1}}
\newcommand{\s}{\sigma}
\renewcommand{\d}{\delta}
\renewcommand{\t}{\theta}
\renewcommand{\l}{\lambda}
\renewcommand{\a}{\alpha}

\newcommand{\rmd}{\mathrm{d}}

\newcommand{\vk}{{\bf{k}}}
\newcommand{\vq}{{\bf{q}}}
\newcommand{\vp}{{\bf{p}}}

\renewcommand{\v}[1]{{\bf{#1}}}
\newcommand{\vs}{\bm{\s}}
\newcommand{\eps}[1]{\epsilon_{#1}}

\newcommand{\su}{\uparrow}
\newcommand{\sd}{\downarrow}

\newcommand{\sx}{\sin k_x}
\newcommand{\sy}{\sin k_y}

\newcommand{\mean}[1]{\big<{#1}\big>}

\newcommand{\etal}{{\emph{et al.}}}
\newcommand{\ie}{\emph{i.e.}~}
\newcommand{\eg}{\emph{e.g.}~}

\usepackage{soul}

\begin{document}

\title{$p$-wave superconductivity in weakly repulsive 2D Hubbard model with Zeeman splitting and weak Rashba spin-orbit coupling}

\author{Henning G. Hugdal}
\email[]{henning.g.hugdal@ntnu.no}
\author{Asle Sudb{\o}}
\email[]{asle.sudbo@ntnu.no}
\affiliation{Department of Physics, NTNU, Norwegian University of Science and Technology, NO-7491 Trondheim, Norway}
\affiliation{Center for Quantum Spintronics, Department of Physics, Norwegian University of Science and Technology, NO-7491 Trondheim, Norway}


\begin{abstract}
We study the superconducting order in a two-dimensional square lattice Hubbard model with weak repulsive interactions, subject to a Zeeman field and weak Rashba spin-orbit interactions. Diagonalizing the non-interacting Hamiltonian leads to two separate bands, and by deriving an effective low-energy interaction we find the mean field gap equations for the superconducting order parameter on the bands. Solving the gap equations just below the critical temperature, we find that superconductivity is caused by Kohn-Luttinger type interaction, while the pairing symmetry of the bands are indirectly affected by the spin-orbit coupling. The dominating attractive momentum channel of the Kohn-Luttinger term depends on the filling fraction $n$ of the system, and it is therefore possible to change the momentum dependence of the order parameter by tuning $n$. Moreover, $n$ also determines which band has the highest critical temperature. Rotating the magnetic field changes the momentum dependence from states that for small momenta reduce to a chiral $p_x \pm i p_y$ type state for out-of-plane fields, to a nodal $p$-wave type state for purely in-plane fields.
\end{abstract}

\maketitle

\section{Introduction}
Much attention has been paid to the possibiblity of unconventional superconductivity due to weak repulsive interactions, as first pointed out by Kohn and Luttinger in 1967.\cite{Kohn1965} They found that due to oscillations in long range interactions, a $p$-wave superconducting state could be formed in a three dimensional electron gas at $\mathcal{O}(U^2)$ in the interaction strength $U$. In two dimensions, no such state can be formed at $\mathcal{O}(U^2)$,\cite{Kagan1989} it is only present at $\mathcal{O}(U^3)$ and zero temperature.\cite{Chubukov1993} However, by applying a magnetic field the effect is present on the majority band also at second order in $U$.\cite{Kagan1989,Raghu2011}

In systems with broken inversion symmetry, either due to crystal structure or an applied electric field, one has to include the effects of spin-orbit interactions by including a Rashba spin-orbit coupling (SOC) term.\cite{Bychkov1984,Manchon2015} A Rashba term in the system Hamiltonian will lead to a coupling between the spin-up and -down Fermi surfaces, and hence opens up the possibility of proximity-induced superconductivity on the minority band.\cite{Raghu2011,Lake2016} Ref.~\onlinecite{Smidman2017} provides a recent review on superconductivity in systems with broken inversion symmetry. The effects of magnetic fields and spin-orbit coupling in two-dimensional systems has been studied in various cases, in limiting cases of \eg the strength of the SOC or the direction of the magnetic field.\cite{Gorkov2001,Barzykin2002,Agterberg2007,Raghu2011,Vafek2011,Raghu2010,Loder2013} Recently Lake \etal \cite{Lake2016} studied a weakly spin-orbit coupled 2DEG with a magnetic field which could be rotated in and out of the plane. They reported that topological $p + ip$ superconductivity is realized when the field is perpendicular to the plane, while an in-plane magnetic field in the $x$-direction leads to a $p_y$ momentum dependence of the order parameter. In either case, only the majority band was found to be superconducting.


In this paper, we perform an analysis similar to that of Ref.~\onlinecite{Lake2016} to study the superconducting order in a weakly repulsive, spin-polarized Hubbard model on a 2D square lattice with weak SOC. Such systems can be realized \eg at the interface between LaAlO$_3$ and SrTiO$_3$, which has been shown to exibit a 2D superconducting state,\cite{Reyren2007,Caviglia2008,Gariglio2009} a magnetic state,\cite{Brinkman2007} and coexistence of superconductivity and magnetism.\cite{Sachs2010,Dikin2011,Bert2011,Li2011} Moreover, it has been shown that the SOC at the interface can be tuned by a gate voltage or an applied electric field.\cite{BenShalom2010,Caviglia2010}

By finding the superconducting state that emerges at the critical temperature $T_c$, we study the dominating pairing symmetries on the two bands for different filling fractions and magnetic field orientations. We find that superconductivity can be induced on \textit{both bands}, depending on the filling fraction. We also find that two different pairing symmetries are realized, one for nearly empty or nearly filled bands, and one close to half filling. However, the small-momentum limit of the order parameters are the same in both regions, a chiral $p_x \pm i p_y$ symmetry for purely out-of-plane fields, and $p$-wave state state for purely in-plane fields. We also find that the Cooper pairs have a finite center-of-mass momentum\cite{Michaeli2012,Loder2013,Lake2016}, \ie a Fulde-Ferrell-Larkin-Ovchinnikov state (FFLO),\cite{Fulde1964,Larkin1965} whenever the magnetic field has an in-plane component.

The remainder of the paper is organized as follows: The model system is presented in Sec.~\ref{sec:model} together with the derivation of the effective Hamiltonian and self-consistent equations for the mean field superconducting gap. The numerical solution strategy is discussed in Sec.~\ref{sec:numerical}, the results of which are presented in Sec.~\ref{sec:results}. Finally, we summarize our results in Sec.~\ref{sec:conclusion}.

\section{Model}\label{sec:model}
\begin{figure}[h!tbp]
\includegraphics[width=0.8\columnwidth]{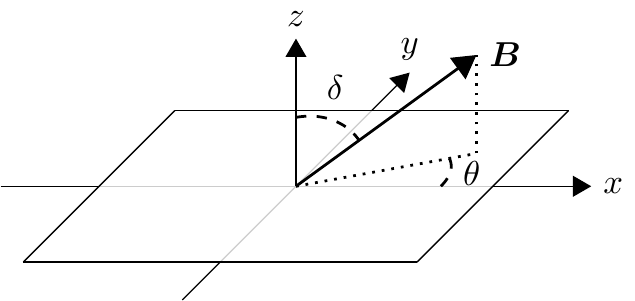}
\caption{\label{fig:geometry}Sketch of system geometry, where the 2D lattice is located in the $xy$-plane and the magnetic field $B$ can point in any direction.}
\end{figure}

Our starting point is a two-dimensional lattice in the presence of an external magnetic field, and with broken inversion symmetry such that SOC is present. A sketch of the geometry is shown in Fig.~\ref{fig:geometry}. We use the Hubbard model augmented by SOC to describe the fermions on the lattice, with a spin-diagonal hopping integral between nearest-neighbor lattice sites given by $t$, and the electrons interact via a on-site repulsion $U n_{i\su} n_{i\sd}$, $U>0$. We will assume that the interaction is weak, \ie the energy scale of the Hubbard-interaction is small compared to the kinetic energy, $U/t\ll 1$.
Time-reversal symmetry is broken by applying an external magnetic field $\v{B}$, which couples to the electrons via the Zeeman coupling $-g\mu_B\v{B}\cdot \vs/2$, where $g$ is the $g$-factor, and $\mu_B$ is the Bohr magneton. This lifts the degeneracy between the spin directions. The effect of SOC is included via a Rashba term with spin-orbit axis normal to the lattice plane, $\alpha_R(\v{p}\times \vs)\cdot \hat{z}$, where $\alpha_R$ is the strength of the spin-orbit coupling. We thus obtain the total system Hamiltonian $H = H_t + H_B + H_{R} + H_I = H_0 + H_I$, with
\begin{subequations}
\begin{eqnarray}
    H_t &=& \sum_{\s, \vk}\eps{\vk} \cre{\vk\s}\anni{\vk\s},\\
    H_{R} &=& \alpha_R \sum_{\vk}\sum_{\s,\s'} (\s^y_{\s\s'}\sin k_x - \s^x_{\s\s'}\sin k_y) \cre{\vk\s}\anni{\vk\s'},\\
    H_B &=& - \v{H}\cdot\sum_{\vk} \sum_{\s,\s'} \vs_{\s\s'} \cre{\vk\s} \anni{\vk\s'},
\end{eqnarray}
\begin{eqnarray}
    H_I &=& \frac{U}{V} \sum_{\vk_1,\vk_2,\vk_3} \cre{\vk_1\su}\cre{\vk_2\sd}\anni{\vk_3\sd}\anni{\vk_1+\vk_2-\vk_3,\su},
\end{eqnarray}
\label{hamiltonian}%
\end{subequations}
where $\eps{\vk} \equiv-2t(\cos k_x + \cos k_y)-\mu$ is the square-lattice tight-binding dispersion relative the chemical potential $\mu$,  $\sigma = \su,\sd$ denotes spin-up and -down electrons respectively, $\v{H} = g\mu_B \v{B}/2 = h(\cos\t \sin\d\hat{x} + \sin\t \sin\d \hat{y} +\cos\d\hat{z})$, and $V$ is the volume of the system. For notational simplicity we have set $\hbar$ and the lattice constant $a$ to $1$ throughout the paper.

\subsection{Diagonalization of non-interacting Hamiltonian}
\begin{figure}[h!tbp]
  \includegraphics[width=\columnwidth]{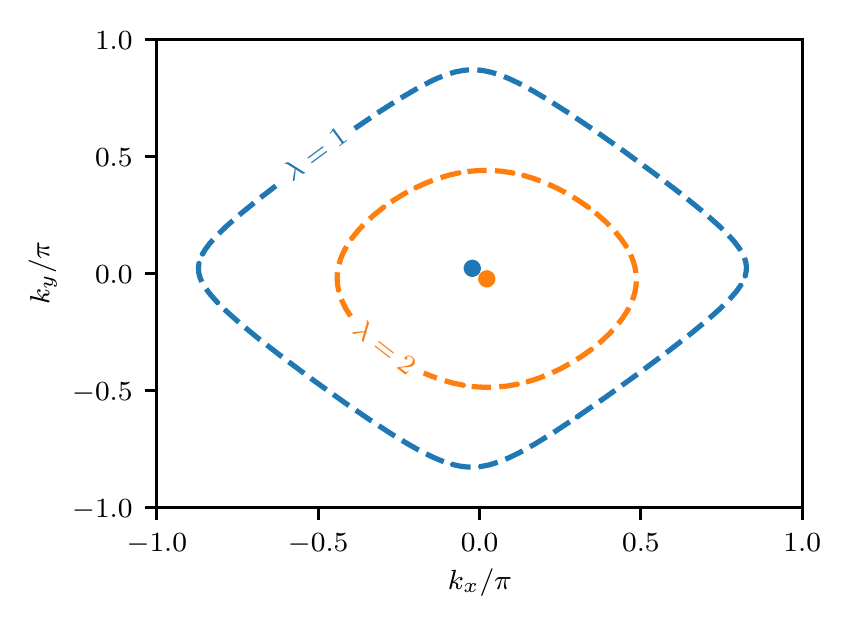}
  \caption{\label{fig:fermi_level}Plot of the Fermi levels for an in-plane magnetic field ($\delta=\pi/2$) with angle $\theta = \pi/4$ relative to the $x$-axis, for filling fraction $n=0.3$, magnetic field strenght $h/t=1$, and SOC strength $\alpha_R/t = 0.2$. The momenta corresponding to the minima of the band dispersions are shifted away from the origin according to Eq.~(\ref{q_shift}). Note that the shift is exaggerated compared to what will be considered throughout the paper.}
\end{figure}

Following Ref. \onlinecite{Lake2016}, we will treat the SOC as a perturbation, assuming that $\alpha_R/h \ll 1$. Hence, we expect that when diagonalizing the non-interacting Hamiltonian $H_0$, the lowest order expression will simply be that of a tight-binding system with spins polarized along the direction of $\v{H}$. We therefore rotate the spin quantization axis to point along the magnetic field using the unitary rotation operator $R_n(\alpha) = \mathrm{exp}({-\i\alpha \vs\cdot\hat{n}/2})$, where $\alpha$ is the angle of rotation about an axis $\hat{n}$: we first rotate an angle $\theta$ about $\hat{n} = \hat{z}$, and then an angle $\delta$ about $\hat{n} = \hat{y}$. This yields
\begin{equation}
  H_0 = \sum_{\vk}\sum_{\sigma,\sigma'} E_{\sigma\sigma'}(\vk)\cre{\vk\sigma}\anni{\vk\sigma'},
\end{equation}
where
\begin{equation}
  \begin{aligned}
    E(\vk) = \eps{\vk}\sigma^0 - h\sigma^z + \alpha_R\big[&(\sx \sin\theta - \sy\cos\theta)\cos\delta\sigma^x\\
     + &(\sx \cos \theta + \sy\sin\theta)\sigma^y\\
    + &(\sy \sin\theta-\sy \cos \theta)\sin\delta \sigma^z\big].
  \end{aligned}
\end{equation}
Diagonalizing $H_0$ leads to two bands with eigenenergies
\begin{widetext}
\begin{equation}
  \begin{aligned}
    \eps{\l}(\vk) &= \eps{\vk} - \zeta_\l\sqrt{h^2 - 2h\a_R(\sx \sin\t - \sy\cos\t)\sin\d + \a_R^2(\sin^2k_x+\sin^2k_y)}\\
    &\approx \eps{\vk} - \zeta_\l \Big[h - \a_R(\sx \sin\t - \sy\cos\t)\sin\d
    + \frac{\a_R^2}{2h}(\sx \cos\t + \sy\sin\t)^2
    + \frac{\a_R^2}{2h}(\sx \sin\t - \sy\cos\t)^2\cos^2\d \Big],\label{eps_approx}
  \end{aligned}
\end{equation}
\end{widetext}
where $\zeta_{\lambda = 1(2)} = +(-) 1$ for the majority (minority) band. In the last line we have kept terms only up to first order in $\alpha_R/h$. In the limit $|\vk| \ll 1$ and $\theta = 0$, this result agrees with Ref.~\onlinecite{Lake2016}. When the magnetic field has an in-plane component, the momentum $\vq$ corresponding the minima of the band dispersions will shift away from the origin according to
\begin{subequations}
  \label{q_shift}
  \begin{eqnarray}
    q_x 
    &\approx& -\frac{\zeta_\l \a_R}{2t}\sin\d\sin\t,\\
    q_y 
    &\approx& +\frac{\zeta_\l \a_R}{2t}\sin\d\cos\t.
  \end{eqnarray}
\end{subequations}
This shift is illustrated in Fig.~\ref{fig:fermi_level}.

Using the eigenvalues in Eq.~(\ref{eps_approx}), we also find relations between the spin and band creation and annihilation operators, which to second order in $\alpha_R/h$ are given by
\begin{subequations}
    \begin{eqnarray}
        \anni{\vk\su} &=&\left[1-\frac{\a_R^2}{8h^2}|\gamma(\vk,\delta,\theta)|^2\right]a_{\vk1} + \frac{\a_R}{2h}\gamma(\vk,\delta,\theta)a_{\vk2}, \\
        \anni{\vk\sd} &=& -\frac{\a_R}{2h}\gamma^\dagger(\vk,\delta,\theta) a_{\vk1}
        + \left[1-\frac{\a_R^2}{8h^2}|\gamma(\vk,\delta,\theta)|^2\right]a_{\vk2},
    \end{eqnarray}
\label{op_transf}
\end{subequations}
where we have defined the function
\begin{equation}
  \begin{aligned}
    \gamma(\vk,\delta,\theta) = &(\sx \sin\t - \sy \cos\t)\cos\d\nonumber \\
      &\qquad- \i(\sx \cos\t + \sy \sin\t)
  \end{aligned}
\end{equation}
and $a^\dagger_{\vk\l}$ and $a_{\vk\l}$ are the creation and annihilation operators for band $\l$ respectively. Using these relations we find that the expectation value of the $z$-component of the spin is $1/2$ for the majority $\lambda=1$ band, and $-1/2$ for the minority $\lambda = 2$ band, with the corrections being second order in $\alpha_R/h$. Hence, to lowest order, the majority and minority bands consist of spin-up and -down particles, respectively. This has consequences for the momentum dependence of any intra-band interaction which could lead to superconductivity. In the next section we will transform the interaction Hamiltonian using the above operator relations and obtain an effective low-energy theory using a Schrieffer-Wolff transformation.\cite{Schrieffer1966}

\subsection{Transformation of the interaction Hamiltonian}
Since the interaction Hamiltonian $H_I$ is proportional to $U$, where we have assumed that the interaction is weak, $U/t \ll 1$, we have to consider what powers of $U$ and $\alpha_R/h$ to keep when transforming the Hamiltonian to the eigenbasis according to Eq.~(\ref{op_transf}). Following Ref.~\onlinecite{Lake2016}, we keep terms of $\mathcal{O}(U^2/t^2)$ and $\mathcal{O}(U\alpha_R^2/th^2)$, while disregarding terms of $\mathcal{O}(U^2\alpha_R/t^2h)$, \ie we assume $\alpha_R/h \gg U/t$.

Transforming the creation and annihilation operators in $H_I$, we get 4 main types of terms: intra-band and pair-hopping terms $\acre{\vk_1\lambda}\acre{\vk_2\lambda}\aanni{\vk_3\mu}\aanni{\vk_4\mu}$ of $\mathcal{O}(U\alpha_R^2/h^2)$, inter-band terms $\acre{\vk_1\lambda}\acre{\vk_2\bar{\lambda}}\aanni{\vk_3\bar{\lambda}}\aanni{\vk_4\lambda}$ of $\mathcal{O}(U)$, and mixed terms such as $\acre{\vk_1\lambda}\acre{\vk_2\lambda}\aanni{\vk_3\lambda}\aanni{\vk_4\bar{\lambda}}$ of $\mathcal{O}(U\alpha_R/h)$ and higher. The notation $\bar{\l}$ denotes the opposite band of $\lambda$.
We collect the intra-band and pair-hopping terms in $H_1$ and the remaining terms in $H_2$:
\begin{equation}
\begin{aligned}
    H_1 &= \sum_{\vk,\vk',\vq}\sum_{\l, \mu} \frac{U\alpha_R^2}{4Vh^2} \Gamma_\l\left(\vk+\frac{\vq}{2}\right)\Gamma_\mu^\dagger\left(\vk'+\frac{\vq}{2}\right) \\
    &\qquad\qquad \times \acre{-\vk+\frac{\vq}{2},\l}\acre{\vk+\frac{\vq}{2},\l}\aanni{\vk'+\frac{\vq}{2},\mu}\aanni{-\vk'+\frac{\vq}{2},\mu} \label{H1},
\end{aligned}
\end{equation}
where
\begin{equation}
\begin{aligned}
  \Gamma_{\l}(\vk) = \zeta_{\l}\big[&\sin k_x \cos\t + \sin k_y \sin\t \\
  &+ i \zeta_{\l}(\sin k_x \sin\t - \sin k_y \cos\t)\cos\d \big],\label{gamma_lambda}
\end{aligned}
\end{equation}
and
\begin{equation}
\begin{aligned}
    H_2 &= \frac{U}{2V} \sum_{\vk_1,\vk_2,\vk_3}\sum_{\l}
    \acre{\vk_1 \l}\acre{\vk_2 \bar{\l}}\aanni{\vk_3 \bar{\l}}\aanni{\vk_1+\vk_2-\vk_3, \l} + \mathcal{O}\left(\frac{U\alpha_R}{h}\right).\label{H2}
\end{aligned}
\end{equation}
The terms in $H_2$ correspond to processes where the resulting quasiparticles are on different bands, and including such interactions in a mean-field treatment would require order parameters with mixed band indices. In order to get a form of the interaction suitable for analysis within a mean-field theory, we perform a Schrieffer-Wolff transformation, see \eg Ref.~\onlinecite{Bravyi2011} for a review. This enables us to get rid of the lowest order processes in $H_2$ while still including the effects of $H_2$ to higher order, such as an intra-band process at $\mathcal{O}(U^2)$. This is
obtained by the unitary transformation
\begin{equation}
\begin{aligned}
  H' = e^{-S}He^{S} = H_0 &+ H_1 + H_2 + [H_0 + H_1 + H_2, S] \\
  &+  \frac{1}{2}[[H_0 + H_1 + H_2, S], S] + ...\label{H_transformed}
\end{aligned}
\end{equation}
where $S$ is an anti-unitary operator chosen such that $[H_0,S] = -H_2$. The lowest order term in $S$ is necessarily of $\mathcal{O}(U/t)$, and this is the only contributing term to the order we are working. Using as an ansatz $S =  \sum_{\vk_1,\vk_2,\vk_3}\sum_\l C_\l(\vk_1,\vk_2,\vk_3,\vk_4)a_{\vk_1\l}^\dagger a_{\vk_2\bar{\l}}^\dagger a_{\vk_3\bar{\l}}a_{\vk_4\l}$, where $\vk_4 = \vk_1+\vk_2-\vk_3$, we find
\begin{equation}
  S = \frac{U}{2V} \sum_{\vk_1, \vk_2, \vk_3,\vk_4} \sum_{\l} \frac{\acre{\vk_1 \l}\acre{\vk_2 \bar{\l}}\aanni{\vk_3 \bar{\l}}\aanni{\vk_4 \l}\delta(\vk_1+\vk_2-\vk_3-\vk_4)}{\eps{\l}(\vk_4)+\eps{\bar{\l}}(\vk_3)-\eps{\bar{\l}}(\vk_2)-\eps{\l}(\vk_1)}.
\end{equation}
Since $S$ comes with a factor $U$, we can neglect most of the terms in the transformed Hamiltonian, leaving us with $H' = H_0 + H_1 + [H_2, S]/2$. Hence, the contributing higher order processes due to $H_2$ are found by calculating the commutator between $H_2$ and $S$.

\begin{figure}[h!tbp]
  \includegraphics[width=0.8\columnwidth]{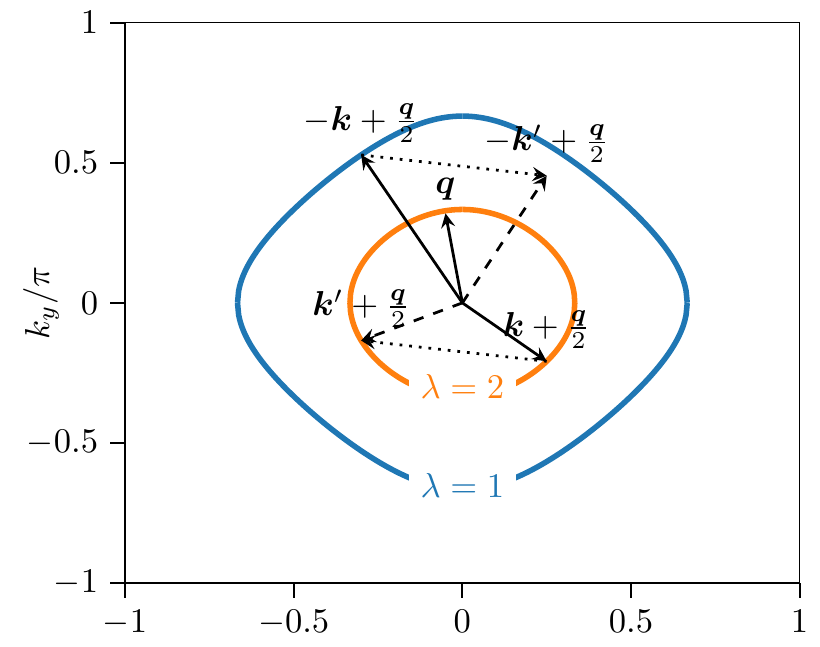}
  \includegraphics[width=0.8\columnwidth]{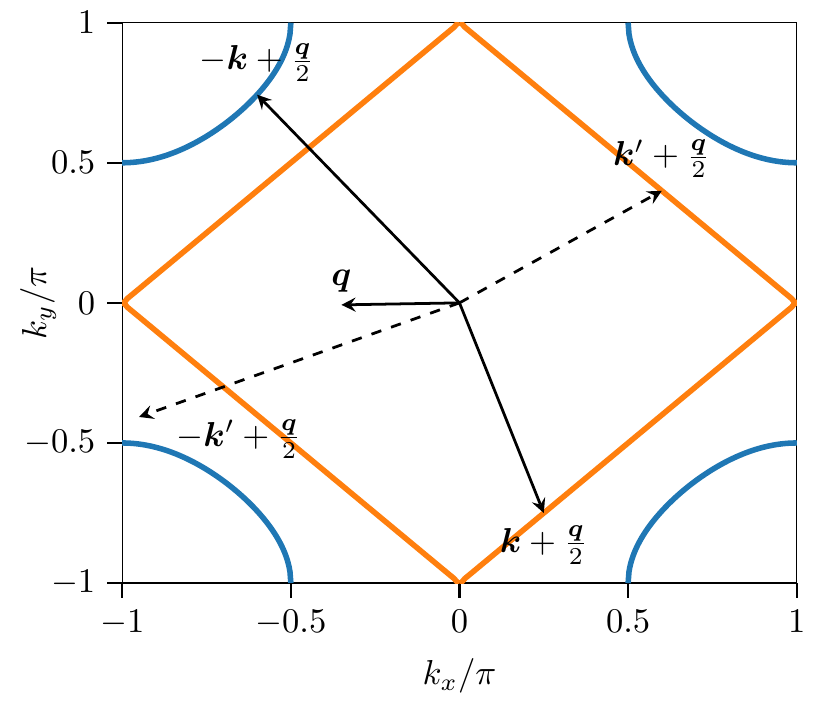}
  \caption{\label{fig:inter_band}The figures illustrate that the inter-band scattering from $\vk +\vq/2$ and $-\vk+\vq/2$ to $\vk' +\vq/2$ and $-\vk'+\vq/2$ has a very limited phase-space for both low (top) and high (bottom) filling fractions $n$. Here we have not included the shifts in center-of-mass momenta, since the shifts are small when $\alpha_R/h \ll 1$.}
\end{figure}

The commutator leads to two kinds of terms of relevant order: a 4-operator inter-band term proportional to $\acre{\l}\acre{\bar{\l}}\aanni{\bar{\l}}\aanni{\l}$ and 6-operator terms $\acre{\l}\acre{\l}\aanni{\l}\aanni{\l}\acre{\bar{\l}}\aanni{\bar{\l}}$, both of $\mathcal{O}(U^2/t^2)$. However, since the interactions must conserve momentum, and the interacting particles lie close to the Fermi level, the phase-space of the inter-band interaction is severely limited, as illustrated in Fig.~\ref{fig:inter_band}. Although the figure does not include the shifts in the minima of the dispersions away from the origin, Eq.~(\ref{q_shift}), these shifts are small when $\alpha_R/h \ll 1$, and the argument should still hold. Hence we will neglect this term, and include only the 6-operator terms.

An effective intra-band process on band $\l$ is obtained from the 6-operator terms $\acre{\l}\acre{\l}\aanni{\l}\aanni{\l}\acre{\bar{\l}}\aanni{\bar{\l}}$ by projecting the operators $\acre{\bar{\l}}\aanni{\bar{\l}}$ to the non-interacting $\bar{\l}$ band, which results in a replacement
$\acre{\bar{\l}\vk}\aanni{\bar{\l}\vk'}\rightarrow \delta(\vk-\vk')f(\eps{\bar{\l}}(\vk))$,\cite{Raghu2011,Lake2016}.  Here, $f(\epsilon)$ is the Fermi-Dirac distribution function. Since the shifts in center-of-mass momenta are small, including them in the interaction terms leads to a correction of higher order than we are considering. We therefore specialize to the case where the total momentum of the particles interacting is zero, which yields the result for the commutator
\begin{equation}
    \begin{aligned}
        \frac{1}{2}[H_2, S] = \frac{U^2}{2V}\sum_{\vk, \vk'}\sum_{\l} \chi_{\bar{\l}}(\vk-\vk')\acre{-\vk',\l}\acre{\vk',\l}\aanni{\vk,\l}\aanni{-\vk,\l},
    \end{aligned}
\end{equation}
where we have defined the susceptibility
\begin{equation}
    \chi_{\l}(\vq) = \frac{1}{V}\sum_{\vp} \frac{f(\eps{\l}(\vp+\vq)) - f(\eps{\l}(\vp))}{\eps{\l}(\vp+\vq)-\eps{\l}(\vp)}.\label{susceptibility}
\end{equation}
In contrast to the 2DEG case,\cite{Raghu2011,Lake2016} we have not been able to calculate the susceptibility analytically for the lattice model. However, a numerical calculation is possible, the results of which will be discussed in Sec.~\ref{sec:susceptibility}.

Setting the total momentum of an interacting pair of particles to zero also in $H_1$, and collecting all terms, we arrive at the effective low-energy Hamiltonian
\begin{equation}
    H' =  H_0 + \sum_{\vk,\vk'} \sum_{\l,\mu}g_{\l\mu}(\vk,\vk')\acre{-\vk,\l}\acre{\vk,\l}\aanni{\vk',\mu}\aanni{-\vk',\mu},
\end{equation}
where we have defined the interaction matrix
\begin{equation}
    g_{\l\mu}(\vk, \vk') = \frac{U^2}{2V}\delta_{\l\mu}\chi_{\bar{\l}}(\vk'-\vk) + \frac{U\a_R^2}{4Vh^2} \Gamma_{\l}(\vk) \Gamma_{\mu}^\dagger(\vk'),\label{g_lambdamu}
\end{equation}
where $\Gamma_\l(\vk)$ is defined in Eq.~(\ref{gamma_lambda}).
The first term in Eq.~(\ref{g_lambdamu}) is an intra-band interaction due to the Kohn-Luttinger mechanism. The second term, which is caused by the SOC, contains both intra-band and pair-hopping terms, with opposite signs due to the factors $\zeta_\l\zeta_\mu$. We thus expect the two terms in Eq.~(\ref{g_lambdamu}) to give rise to different superconducting states. The first term gives rise to uncoupled ordered states on the two bands with different $T_c$, while the second term couples the order parameters and should lead to simultaneous superconductivity on both bands.

\begin{figure*}
    \includegraphics[width=0.4\textwidth]{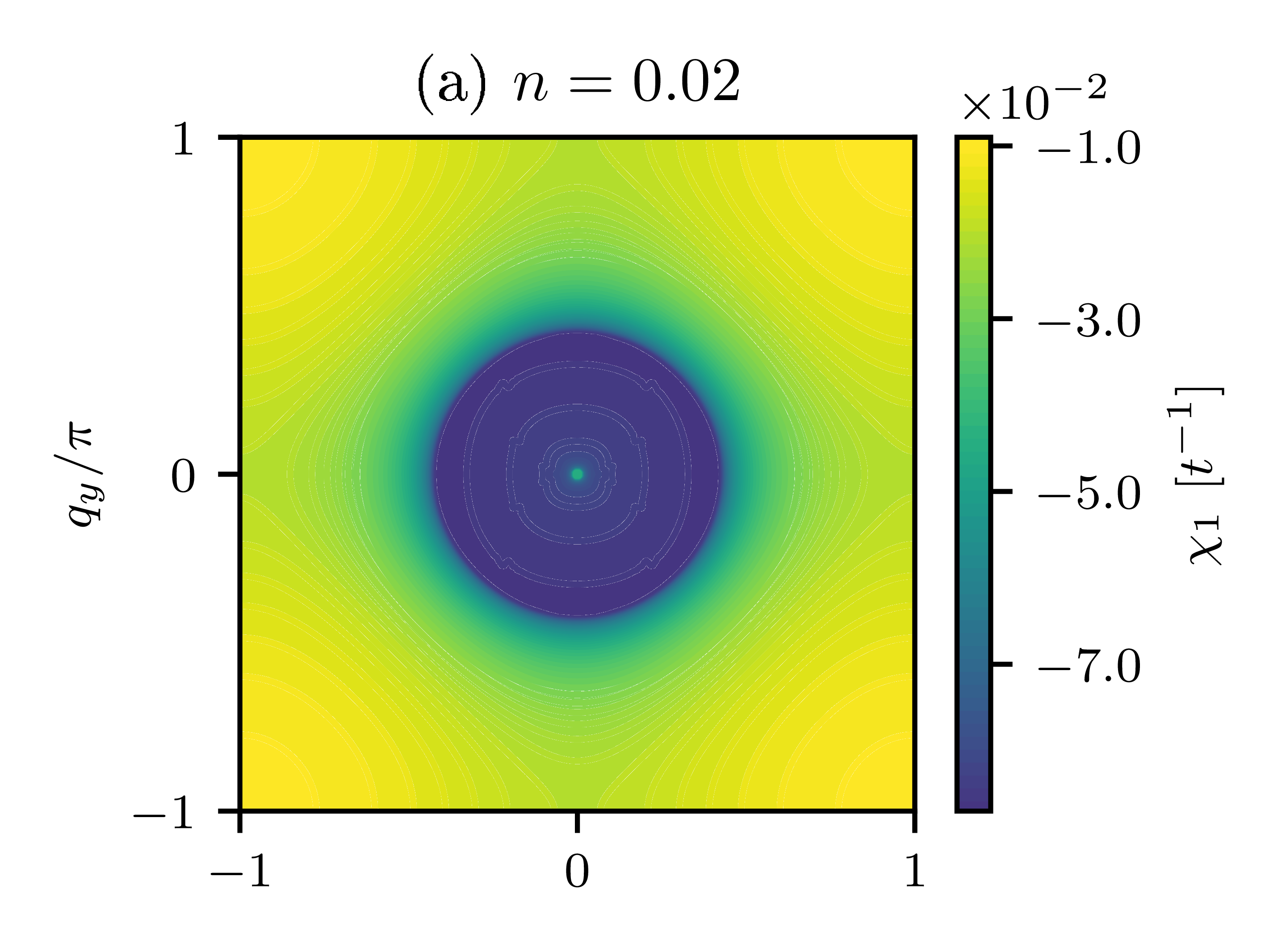}
    \includegraphics[width=0.378\textwidth]{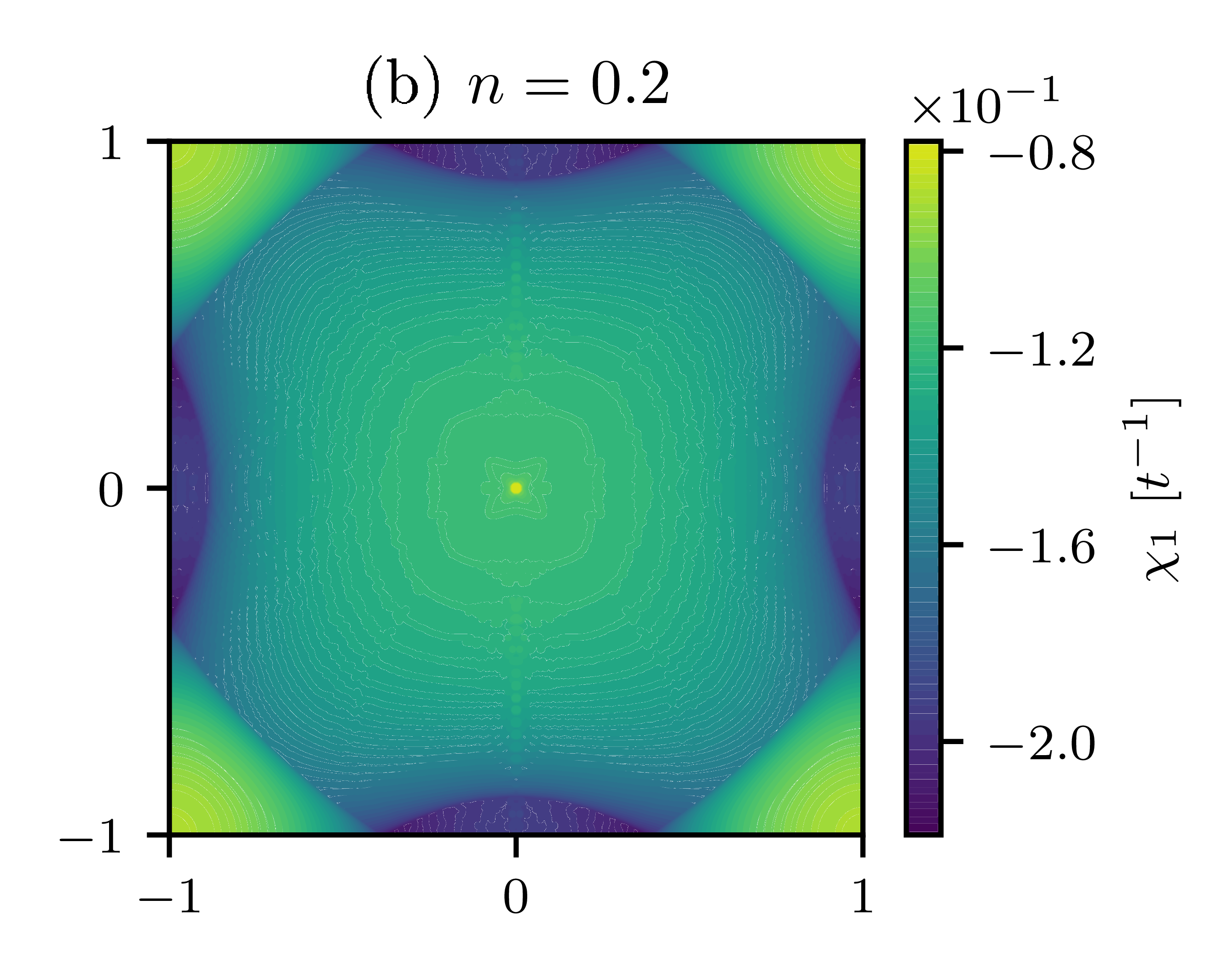}
    \includegraphics[width=0.4\textwidth]{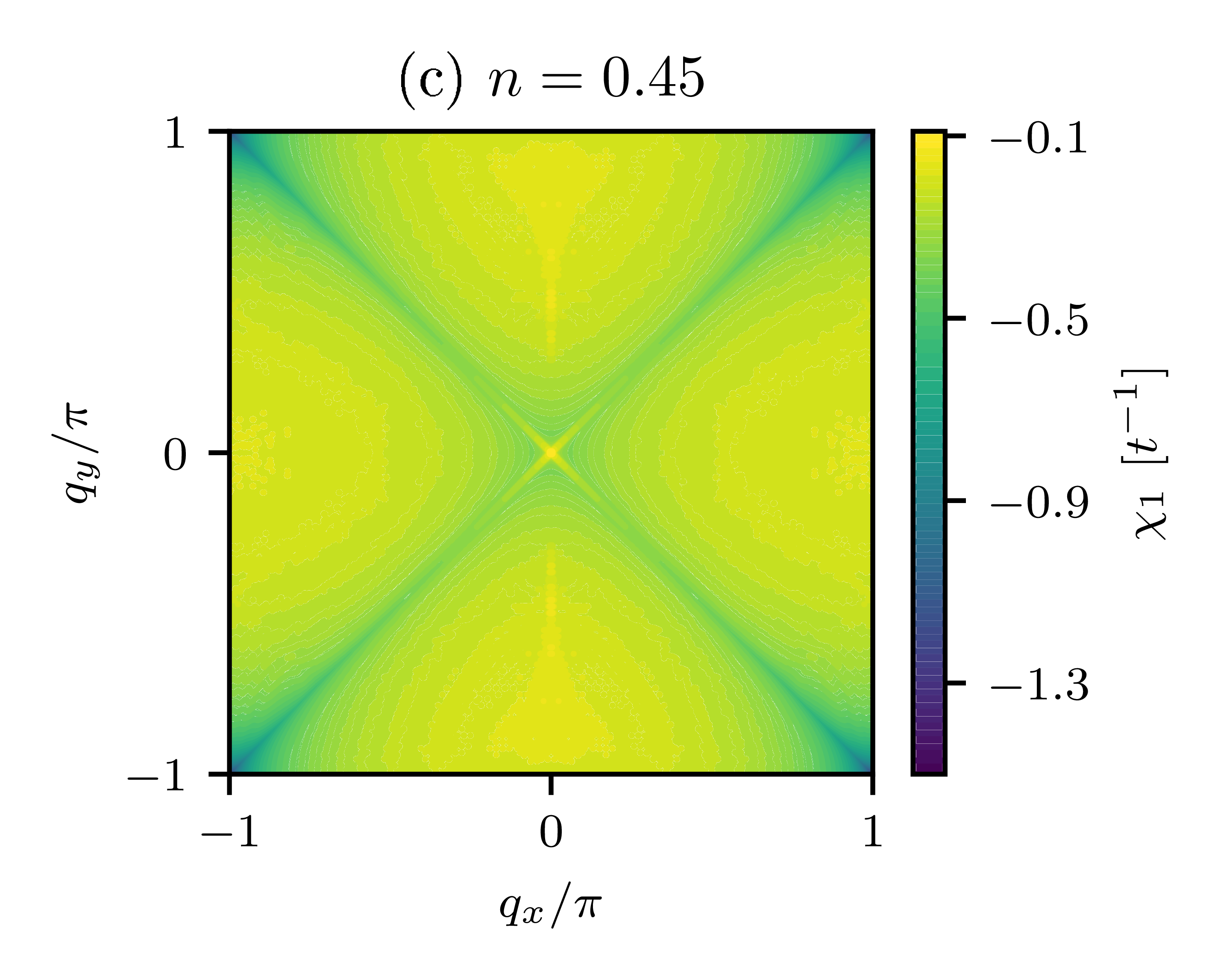}
    \includegraphics[width=0.378\textwidth]{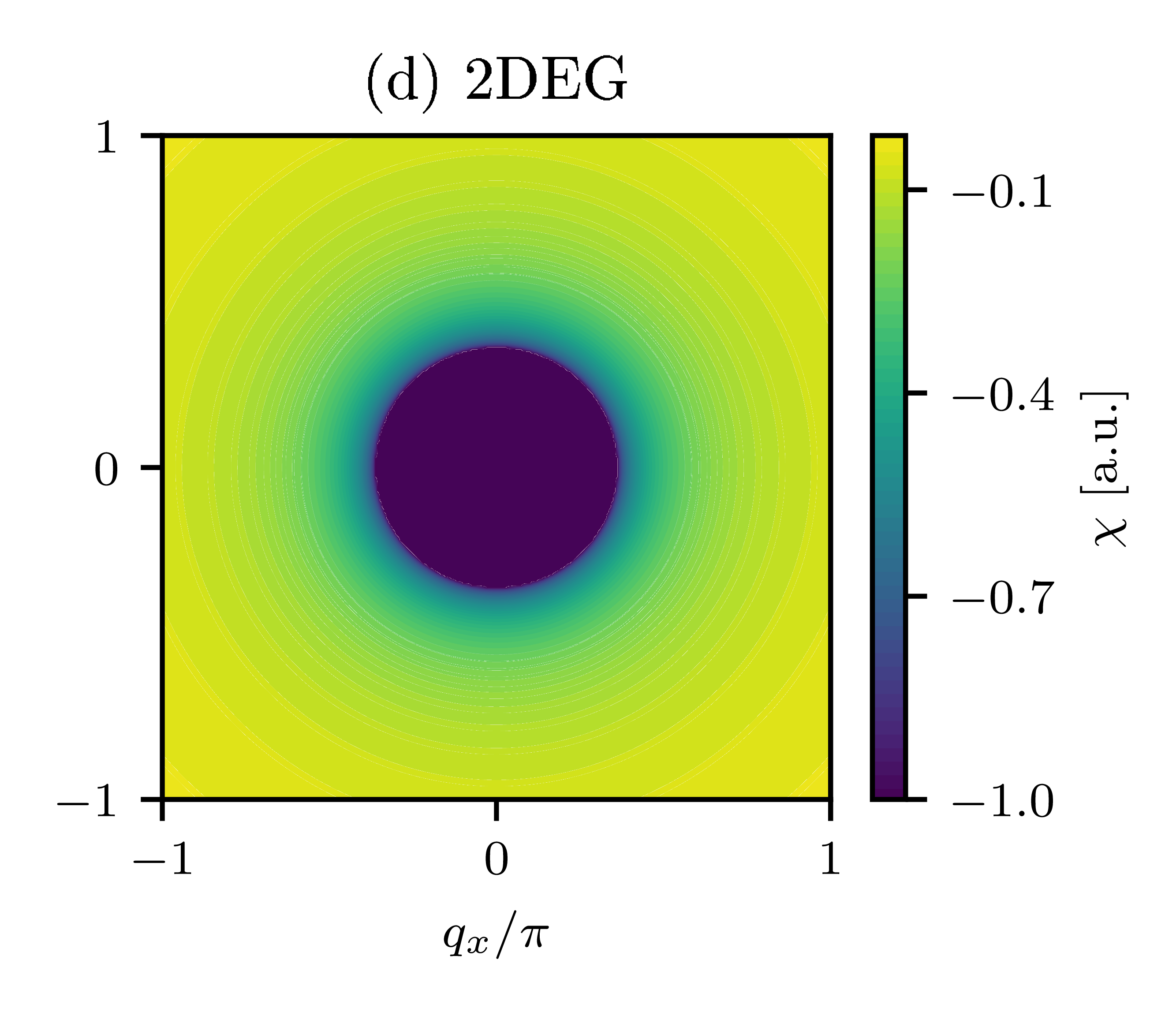}
    \caption{\label{fig:susceptibilities} Plot of numerically calculated susceptibilities for the majority band at filling fraction (a) $n=0.02$, (b) $n=0.2$ and (c) $n=0.45$, which is close to half-filling of the band. The spikes at $\vq=0$ are numerical divergences that do not contribute to the results when expanding in square lattice harmonics. The susceptibility in the 2DEG case with Zeeman splitting treated in Refs.~\onlinecite{Raghu2011,Lake2016}, $\chi(\vq) \propto -1 + \mathrm{Re}\sqrt{q^2-(2k_F)^2}/q$, is shown in (d) with $k_F/\pi = 0.2$ for comparison.}
\end{figure*}

\subsection{Mean field treatment}
Defining the mean-field order parameters (gap-functions)
\begin{eqnarray}
    \Delta_{\l}(\vk) &=& -\sum_{\vk',\mu} 2g_{\l\mu}(\vk,\vk') \mean{\aanni{\vk'\mu}\aanni{-\vk'\mu}},\\
    \Delta_{\mu}^\dagger(\vk) &=& -\sum_{\vk',\l} 2g_{\l\mu}(\vk',\vk)\mean{\acre{-\vk'\l}\acre{\vk'\l}},
\end{eqnarray}
we rewrite the Hamiltonian in the standard way
\begin{eqnarray}
    H' &=&
    \sum_{\vk,\l} \frac{1}{2}\left[(\eps{\l}(-\vk)-\mu) + \Delta_{\l}(\vk)\mean{\acre{-\vk\l}\acre{\vk\l}}\right] \nonumber\\
    &&\qquad\qquad\qquad\qquad + \frac{1}{2}\sum_{\vk,\l} \psi_{\vk\l}^\dagger \mathcal{E}_{\l}(\vk) \psi_{\vk\l}.
\end{eqnarray}
Here, we have defined the Nambu spinors $\psi_{\vk\l} =  (\aanni{\vk\l} \quad \acre{-\vk\l})^T$ and the matrix
\begin{equation}
    \mathcal{E}_{\l}(\vk) = \left(\begin{matrix}
        \eps{\l}(\vk)-\mu & \Delta_{\l}(\vk)\\
        \Delta_{\l}^\dagger(\vk) & -\eps{\l}(-\vk)+\mu
    \end{matrix}\right).
\end{equation}
Performing a Bogoliuobov transformation yields
\begin{equation}
  H' = E_0 + \sum_{\vk,\l} \left[\frac{\eps{\l}(\vk) - \eps{\l}(-\vk)}{2} + E_\l(\vk)\right]n_{\vk\l}
\end{equation}
where $n_{\vk\l}$ is the number operator of the Bogoliubov quasipartices in the rotated basis
\begin{equation}
  E_\l(\vk) = \sqrt{\xi_\l^2(\vk)+|\Delta_\l(\vk)|^2},
\end{equation}
is the approximate quasiparticle dispersion with $\xi_\l(\vk) \equiv (\eps{\l}(\vk) + \eps{\l}(-\vk))/2$. Moreover
\begin{equation}
  E_0 = \frac{1}{2} \sum_{\vk,\l} [\xi_\l(\vk) - E_\l(\vk) + \Delta_\l(\vk)\mean{\acre{-\vk\l}\acre{\vk\l}}].
\end{equation}
Since the SOC term in the system Hamiltonian is the only term which breaks inversion symmetry, we have $(\eps{\l}(\vk) - \eps{\l}(-\vk))/2 \sim \alpha_R$. The $(\eps{\l}(\vk) - \eps{\l}(-\vk))$ term in the diagonalized Hamiltonian thus leads to higher order corrections, and will therefore be neglected. Minimizing the free energy with respect to $\Delta_\lambda^\dagger(\vk)$ yields the gap equations
\begin{equation}
    \Delta_{\l}(\vk) = - \sum_{\vk', \mu} \frac{g_{\l\mu}(\vk,\vk')\Delta_{\mu}(\vk')}{E_\mu(\vk')}\tanh\left(\frac{\beta E_\mu(\vk')}{2}\right),\label{gap_eqs}
\end{equation}
where $\beta = 1/k_B T$. Note that we have set $\alpha_R = 0$ in $E_\l(\vk)$, since including the effects of the SOC in the dispersion give rise to terms of higher order than what we are considering.

\section{Numerical solution strategy}\label{sec:numerical}
\subsection{Calculation of the susceptibility}\label{sec:susceptibility}
The susceptibility is obtained numerically from Eq.~(\ref{susceptibility}) in the zero temperature limit. Since the susceptibility enters the gap equations Eq.~(\ref{gap_eqs}) with a prefactor proportional to  $U^2$, we can neglect the effects of SOC and thus set $\alpha_R=0$ in the calculations. The results for the majority band for three different $n$ are shown in Fig.~\ref{fig:susceptibilities}, together with the analytical result for the 2DEG with Zeeman splitting treated in Refs.~\onlinecite{Raghu2011,Lake2016}. For low $n$ the susceptibility is isotropic, and resembles the 2DEG result. Closer to half-filling the susceptibility becomes more anisotropic due to the anisotropy of the dispersion.

In order to find the dominating attractive pairing channels due to the Kohn-Luttinger term in the gap equations, we expand the results for the susceptibility in square lattice harmonics, see the \hyperref[sec:appendix]{Appendix} for details. Considering only the dominant attractive pairing channels, we find that the susceptibility to good approximation can be written
\begin{equation}
  \begin{aligned}
  \chi_\l(\vk-\vk') &= \chi_\l^1 \big[g_{x+iy}(\vk)g_{x-iy}(\vk') + g_{x-iy}(\vk)g_{x+iy}(\vk')\big]\\
   &+ \chi_\l^2\big[g_{x}(\vk)g_{x}(\vk') + g_{y}(\vk)g_{y}(\vk')\big]\\
   &+ \chi_\l^3\big[g_x(k_x,2k_y)g_x(k_x',2k_y') + g_y(k_x,2k_y)g_y(k_x',2k_y')\\
   &+ g_x(2k_x,k_y)g_x(2k_x',k_y') + g_y(2k_x,k_y)g_y(2k_x',k_y')\big]
   ,\label{chi_expansion}
  \end{aligned}
\end{equation}
where we have defined the functions
\begin{subequations}\label{expansion_functions}%
    \begin{eqnarray}
        2\pi g_{x+iy}(\vk) &=& \sin k_x + i \sin k_y,\label{xpiy}\\
        2\pi g_{x-iy}(\vk) &=& \sin k_x - i \sin k_y,\label{xmiy}\\
        2\pi g_x(\vk) &=& 2\pi g_x(k_x,k_y) = 2\sin k_x \cos k_y, \label{x}\\
        2\pi g_y(\vk) &=& 2\pi g_y(k_x,k_y) = 2\cos k_x \sin k_y\label{y}.
    \end{eqnarray}
\end{subequations}
These functions are orthonormal, \ie $\int_{1\mathrm{BZ}} \rmd \vk ~ g_i(\vk)g_j^\dagger(\vk) = \delta_{ij}$.

\begin{figure}
 \includegraphics[width=\columnwidth]{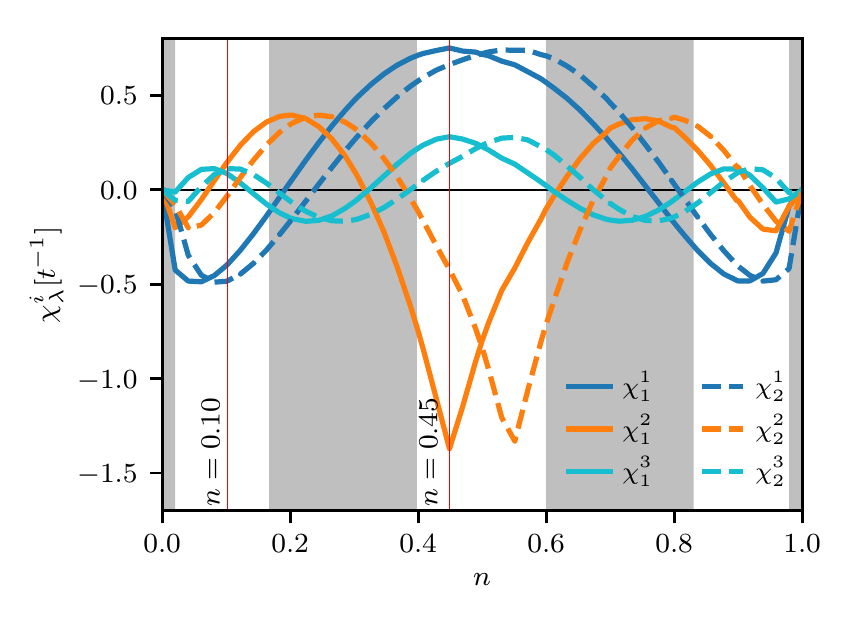}
 \caption{\label{fig:Ds} Plot of expansion coefficients $\chi_\l^i$ in Eq.~(\ref{chi_expansion}) as a function of filling fraction $n$ for $h=0.2t$ at zero temperature. Filling fraction $n=0$ corresponds to a completely empty system, and $n=1$ to two completely filled bands. The gray regions indicate where keeping only the first two terms in the expansion in Eq.~(\ref{chi_expansion}) is not sufficient due to dominant contributions to attractive pairing from other square lattice harmonics, such as the $\chi_\l^3$-term in Eq.~(\ref{chi_expansion}).}
\end{figure}

The values for the expansion coefficients $\chi_\l^i$ for different filling fractions $n$ are shown in Fig.~\ref{fig:Ds} for $h=0.2t$ at zero temperature. Notice that $\chi_\l^i(n) = \chi_{\bar{\l}}^i(1-n)$. We will in the following focus on filling fractions where the first two terms in Eq.~(\ref{chi_expansion}) suffice to describe the most attractive pairing channel, \ie the channel with the most negative coefficient $\chi_\l^i$. Regions where this does not simultaneously hold for both susceptibilities, because of significant or dominant contributions from other channels, are indicated by the gray regions in the figure. The coefficients $\chi_\l^i$ should, strictly speaking, be calculated at the temperature of the system, but we expect the superconducting transition temperature to be sufficiently low for this to be a good approximation.

The plots of the coefficients $\chi_\l^i$ in Fig.~\ref{fig:Ds} illustrate two important points. Firstly, the dominant attractive Kohn-Luttinger pairing channel depends strongly on the filling fraction. For instance the dominant attractive channel for intermediate filling fractions differs from low and high filling fractions. This is related to the shape of the Fermi surfaces in these regions, and could lead to significantly different $\vk$-dependences of the order parameters in these regions.
Secondly, the plots also show that the majority and minority bands have the most negative expansion coefficient in different filling fraction intervals. Therefore, there exists a possibility that there can be a switching between bands with the highest $T_c$.

\subsection{Momentum dependence of the order parameter}
From the preceding subsection, we found that the potentially dominating momentum dependence of the superconducting gap due to the Kohn-Luttinger term in the interaction Eq.~(\ref{g_lambdamu}) could be any of the four functions in Eq.~(\ref{expansion_functions}). If however, the solution were to be determined by the second term in Eq.~(\ref{g_lambdamu}), the solution should be proportional to $\Gamma_\l(\vk)$ in Eq.~(\ref{gamma_lambda}), which can be rewritten in terms of $g_{x\pm iy}(\vk)$,
\begin{equation}
  \begin{aligned}
    \Gamma_{\l}(\vk) = \pi& g_{x+iy}(\vk)(\xi_{\l} - \cos\d)(\cos\t- i\sin\t)\\
    &+ \pi g_{x-iy}(\vk)(\xi_{\l}+\cos\d)(\cos\t+i\sin\t).
\end{aligned}
\end{equation}
Therefore, keeping only the dominant terms, the superconducting gap can be expanded using the four functions in Eq.~(\ref{expansion_functions}),
\begin{equation}
    \Delta_\l(\vk) = \Delta_\l^{x+iy}g_{x+iy}(\vk) + \Delta_\l^{x-iy}g_{x-iy}(\vk) +\Delta_\l^{x}g_{x}(\vk) + \Delta_\l^{y}g_{y}(\vk).
    \label{Delta_expansion}
\end{equation}

\subsection{Solutions close to the critical temperature $T_c$}
The physically realizable solution of the gap equations is the solution which corresponds to a global minimum of the free energy. However, when solving the gap equations numerically using \eg a root solver, the solution might just as well correspond to a local minimum of the free energy. These solutions will have a lower $T_c$, and will therefore not be realized when cooling down the system. In order to circumvent this problem, we instead calculate $T_c$ and find the corresponding solution.

\begin{figure*}[h!btp]
  \includegraphics[width=\textwidth]{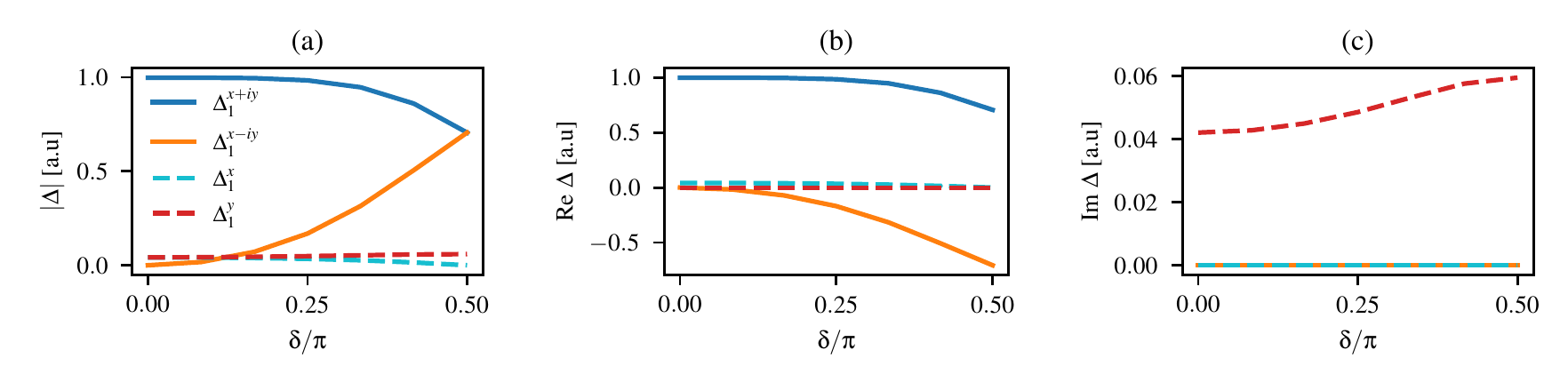}
  \caption{\label{fig:Delta_n01}Plot of the (a) absolute value, (b) real and (c) imaginary part of the dominant elements of the eigenvector of $\mathcal{M}(T_c)$ corresponding to eigenvalue 1 as a function of $\delta$ for $n=0.1$ and  $\theta=0$. The terms proportional to the function $g_{x\pm iy}(\vk)$ are the dominant terms in $\Delta_1(\vk)$. $\Delta_2(\vk) = 0$, not shown in the plot.}
\end{figure*}

Close to and below $T_c$, we linearize the gap equations,
\begin{equation}
  \Delta_{\l}(\vk, T_c^-) = - \sum_{\vk', \mu} \frac{g_{\l\mu}(\vk,\vk')\Delta_{\mu}(\vk', T_c^-)}{|\xi_\mu(\vk')|}\tanh\left(\frac{\beta_c |\xi_\mu(\vk')|}{2}\right).
\end{equation}
By multiplying this equation by $(2\pi)^2g_j^\dagger(\vk)/V$, where $j=\{x+iy,x-iy,x,y\}$, and summing over the first Brillouin zone, we get a system of linear equations
\begin{equation}
  \Delta_\l^i = \sum_{j} \sum_{\mu} \mathcal{M}_{\l\mu}^{ij}(T_c) \Delta_\mu^j,
\end{equation}
where
\begin{equation}
  \begin{aligned}
    \mathcal{M}_{\l\mu}^{ij}(T_c)
     = - \frac{(2\pi)^2}{V}\sum_{\vk'} \Bigg[&\left(\sum_{\vk} g_{\l\mu}(\vk,\vk')g_j^\dagger(\vk)\right)\\
    & \times \frac{g_i(\vk')}{|\xi_\mu(\vk')|}\tanh\left(\frac{\beta_c |\xi_\mu(\vk')|}{2}\right)\Bigg].
  \end{aligned}
\end{equation}
which may conveniently be written on the form
\begin{equation}
  \vec{\Delta} = \mathcal{M}(T_c)\vec{\Delta}.
\end{equation}
Here, $\vec{\Delta} = (\Delta_1^{x+iy}\quad \Delta_1^{x-iy}\quad \dots\quad \Delta_2^{y})^T$.
Thus, for a non-trivial solution to exist we require that $\mathrm{det}(\mathcal{M}(T_c)) = 0$, which allows for a computation of $T_c$. In cases where this holds for multiple temperatures, the highest $T_c$ corresponds to the channel where superconductivity actually occurs. When $T_c$ is determined, $\vec{\Delta}$ is found by calculating the eigenvector of $\mathcal{M}(T_c)$ corresponding to eigenvalue $1$. The eigenvector only gives information about the relative size of the coefficients in Eq.~(\ref{Delta_expansion}), not the absolute scale. This is nonetheless enough information to determine the dominant momentum dependence of the order parameter close to $T_c$, and hence in which channel superconductivity first appears upon cooling.

\section{Results and discussion}\label{sec:results}
Using the procedure described in the previous section, we have calculated the eigenvector of $\mathcal{M}(T_c)$, focusing on filling fractions $n=0.1$ and $n=0.45$. These values are indicated in Fig.~\ref{fig:Ds}. All results are obtained with $h=0.2t$. For $n=0.1$, the results as a function of tilt angle $\delta$ at $\theta=0$ is shown in Fig.~\ref{fig:Delta_n01}.
We see that for a pure out-of-plane field, $\delta=0$, $\Delta_1(\vk) \propto \sin k_x + i\sin k_y$, which for small momenta corresponds to a chiral $k_x + i k_y$ order parameter. For a pure in-plane field in the $x$-direction, $\Delta_1(\vk) \propto \sin k_y$ which corresponds to a $k_y$-dependence in the low $|\vk|$ limit. This is in agreement with the results of Lake \etal\cite{Lake2016} It is important to note that when calculating the eigenvectors at $T_c$, we do not get information about the absolute value of the gaps, nor the relative size of the gap coefficients between \eg $\delta=0$ and $\delta=\pi$.

\begin{figure}
    \includegraphics[width=0.9\columnwidth]{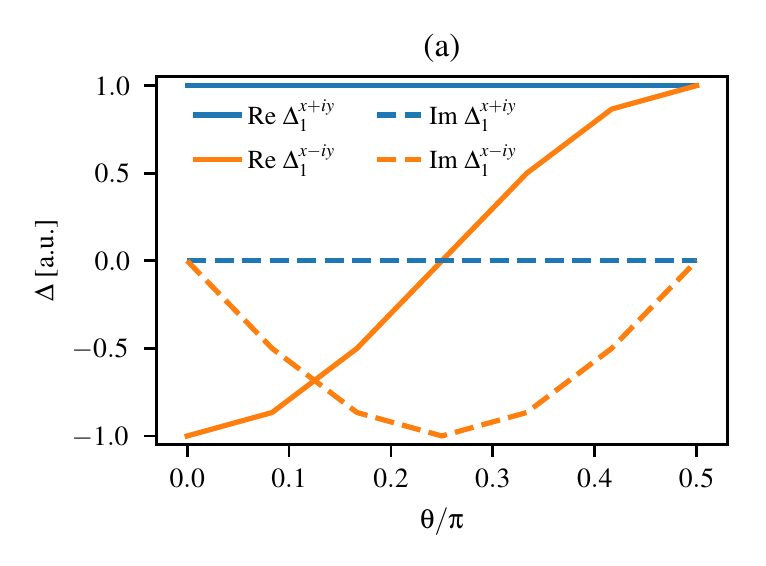}
    \includegraphics[width=0.9\columnwidth]{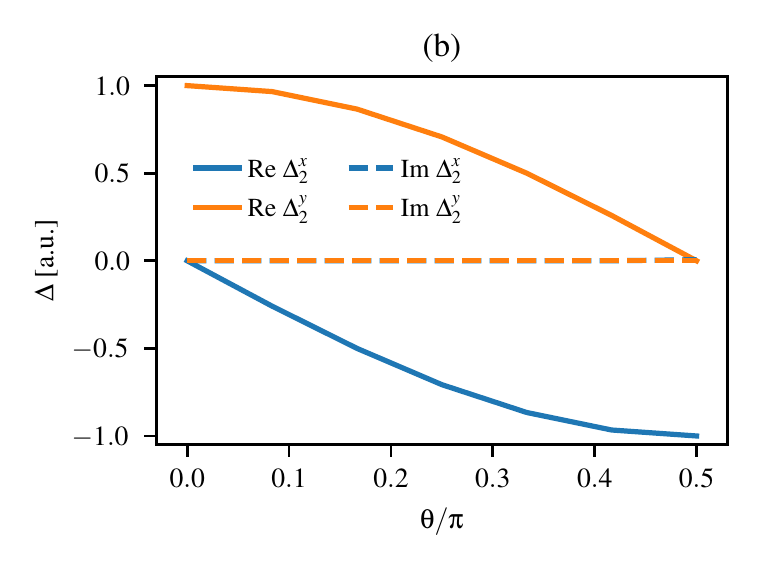}
    \caption{\label{fig:Delta_theta} Plot of dominating terms of the eigenvector as a function of $\theta$ for pure in-plane magnetic field and filling fraction (a) $n=0.1$ and (b) $n=0.45$. In the small-$|\vk|$ limit the $\vk$-dependence is changed from pure $k_y$ to a pure $k_x$ as the field is rotated. The overall phase is chosen such that the dominating contribution at $\theta=0$ is real.}
\end{figure}

Rotating the magnetic field in the $xy$-plane, the $\vk$-dependence of the gap also changes accordingly, from a pure $\sin k_y$-dependence for $\theta=0$, to a pure $\sin k_x$-dependence for $\theta=\pi/2$, as seen from the values of the coefficients in Fig.~\ref{fig:Delta_theta}(a). 
This change coincides with the rotation of the center momentum $\vq$ in Eq.~(\ref{q_shift}). The reason for this might be that the superconducting state is of FFLO kind whenever there is an in-plane component of the field. In the above calculations, we neglected the shift in the center momentum of the Fermi levels, Eq.~(\ref{q_shift}), since they lead to higher order corrections. However, since the Fermi levels in fact are shifted, the Cooper pairs have a finite center momentum $2\vq$ and thus are FFLO Cooper pairs. This is in agreement with Ref.~\onlinecite{Lake2016}.
\begin{figure*}
  \includegraphics[width=\textwidth]{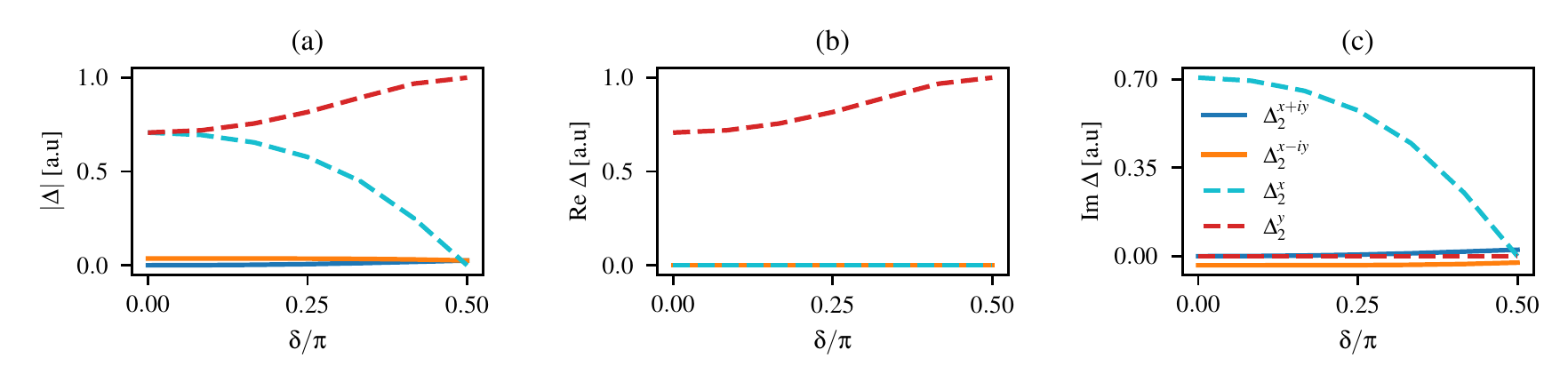}
  \caption{\label{fig:Delta_n045}Plot of the (a) absolute value, (b) real and (c) imaginary part of the dominant elements of the eigenvector of $\mathcal{M}(T_c)$ corresponding to eigenvalue 1 as a function of $\delta$ for $n=0.45$ and  $\theta=0$. The terms proportional to the functions $g_{x}(\vk)$ and $g_{y}(\vk)$ are the dominant terms in $\Delta_2(\vk)$, which differs from the $n=0.1$ case in Fig.~\ref{fig:Delta_n01}. $\Delta_1(\vk) = 0$, not shown in the plot.}
\end{figure*}

Though the majority band has the highest $T_c$ here, we see from Fig.~\ref{fig:Ds} that also the minority band is attractive in the $g_{x\pm iy}$ channel for low filling fractions, in contrast to what has been found in other studies with quadratic dispersions.\cite{Raghu2011,Lake2016} Instead of being completely flat for $|\vk-\vk'| < 2|\vk_{F\l}|$, as in the quadratic case, the susceptibility developes a dome in this region when increasing the filling fraction. In this way, the susceptibility on the majority band also becomes $\vk$-dependent for interactions between particles close to the Fermi surface on the minority band, leading to the possibility of attractive interactions. We therefore expect that the minority band becomes superconducting at some finite temperature lower than $T_c$ on the majority band.

Performing similar calculations close to half-filling, with $n=0.45$, we find that the momentum dependence for the order parameter is dominated by the functions $g_x(\vk)$ and $g_y(\vk)$. Moreover, superconductivity is now induced on the minority band at $T_c$, as shown in Fig.~\ref{fig:Delta_n045}. The value $n=0.45$ is close to the filling fraction for which the majority band is half-filled, which corresponds to a van Hove-singularity in the density of states of the majority band. The fact that the minority band has the highest $T_c$ can thus be explained by the vast number of particles on the majority band which can mediate an effective intra-band interaction. Again the functional form of the gap is changed by rotating the magnetic field: when $\delta=0$, $\Delta_2(\vk)\propto \sin k_x \cos k_y - i\cos k_x \sin k_y$, which in the small-$|\vk|$ limit corresponds to $k_x - ik_y$, and thus has the opposite chirality compared to the $n=0.1$ case. For a pure in-plane field we get $\Delta_2(\vk) \propto \cos k_x \sin k_y$, which for small momenta corresponds to a pure $k_y$-dependence. As for $n=0.1$, rotating the field in-plane changes the $\vk$-dependence, as shown in Fig.~\ref{fig:Delta_theta}(b).

Since the coefficients $\chi_\l^i$ have the symmetry $\chi_\l^i(n) = \chi_{\bar{\l}}^i(1-n)$, we have also performed the above analysis for $n=0.9$ and $n=0.55$. In both cases, superconductivity is now present on the opposite band compared to the $n=0.1$ and $n=0.45$ cases, again with helicity $k_x + ik_y$ for $\l = 1$, and $k_x - i k_y$ for $\l=2$. Therefore, it appears that a superconducting state with the same helicity as the band is favoured.\cite{Lake2016}

In both the previous cases, only one band is superconducting at $T_c$. This indicates that the second term in Eq.~(\ref{g_lambdamu}) does not contribute significantly to the superconducting pairing, as this would lead to simultaneous superconductivity on both bands. Moreover, from the form of $\Gamma_\l(\vk)$, we see that this term should lead to superconductivity of opposite chirality of what was found here, $k_x \mp ik_y$ on the majority/minority band.\cite{Lake2016} Notice also that it is in principle possible to read off the dominating functional form of the superconducting gap directly from Fig.~\ref{fig:Ds}.

Finally, we have found that the value of $\alpha_R/h$ has no impact on $T_c$, while it depends strongly on the value of $U/t$. These are indications that the Kohn-Luttinger term in the interaction is responsible for the physically realizable superconducting order, and thus due to pure intra-band interactions. This allows to make some predictions regarding parts of the gray regions in Fig.~\ref{fig:Ds}, where the $\chi_\l^3$ term is the dominating attractive term. From the above results it is reasonable to assume that the solution in these regions is of the form $\Delta_\l(\vk) = \Delta_\l^{x,2y}g_x(k_x,2k_y) + \Delta_\l^{y,2y}g_y(k_x,2k_y) + \Delta_\l^{x,2x}g_x(2k_x,k_y) + \Delta_\l^{y,2x}g_y(2k_x,k_y)$, with the same small-$|\vk|$ functional form as found above. This has however not been checked explicitly.

The fact that superconductivity is not proximity-induced on the opposite band by the second term in Eq.~(\ref{g_lambdamu}), requires that $\Delta_\l(\vk)$ satisfies
\begin{equation}
  \sum_\vk \frac{\Gamma_\l^\dagger(\vk)\Delta_\l(\vk)}{|\xi_\l(\vk)|}\tanh\left(\frac{\beta|\xi_\l(\vk)|}{2}\right) = 0.
\end{equation}
Using this requirement we derive an ansatz for the functional form of the superconducting gaps,
\begin{subequations}
\begin{equation}
\begin{aligned}
    \Delta_1(\vk) = \Delta_1^1 &\Bigg[\frac{1+\cos\d}{\sqrt{2(1 + \cos^2\delta)}}(\cos\t-i\sin\t)g_{x+iy}(\vk)\\
     &- \frac{1-\cos\d}{\sqrt{2(1 + \cos^2\delta)}}(\cos\t+i\sin\t)g_{x-iy}(\vk)\Bigg]\\
    +  \Delta_1^2 & \Bigg[\frac{\cos\t-i\cos\d\sin\t}{\sqrt{1 + \cos^2\delta}}g_y(\vk) \\
    &\qquad- \frac{\sin\t + i \cos\d\cos\t}{\sqrt{1 + \cos^2\delta}}g_x(\vk)\Bigg],
\end{aligned}
\end{equation}
\begin{equation}
\begin{aligned}
   \Delta_2(\vk) = \Delta_2^1 & \Bigg[\frac{1+\cos\d}{\sqrt{2(1 + \cos^2\delta)}}(\cos\t+i\sin\t)g_{x-iy}(\vk) \\
   &- \frac{1-\cos\d}{\sqrt{2(1 + \cos^2\delta)}}(\cos\t-i\sin\t)g_{x+iy}(\vk)\Bigg]\\
   + \Delta_2^2 & \Bigg[\frac{\cos\t+i\cos\d\sin\t}{\sqrt{1 + \cos^2\delta}}g_y(\vk) \\
   &\qquad- \frac{\sin\t - i \cos\d\cos\t}{\sqrt{1 + \cos^2\delta}}g_x(\vk)\Bigg],
\end{aligned}
\end{equation}
\end{subequations}
where $\Delta_\l^i$ in general can depend on the field alignment angle.
Using this ansatz to find $T_c$ and the solution eigenvectors, we find the same results as presented above. Hence, we see that even though the results do not depend directly on the SOC strength, the fact that SOC is present affects the realized pairing symmetry.\cite{Lake2016} The results of Ref.~\onlinecite{Loder2013} indicate that this conclusion might not hold for all values of $\alpha_R/h$, and an interesting development would therefore be to study this system for general SOC strengths.

The $T_c$ quickly decreases with decreasing $U/t$, and for values in the regime set by the derivation of the gap equations, a numerical solution is impossible. Hence, we have performed the above analysis for a range of values of $U/t$ and $\alpha_R/h$, and found that the results were qualitatively unchanged. The fact that the results agree with Ref.~\onlinecite{Lake2016} for small filling fractions, and that $T_c$ depends only on $U/t$ indicate that the results presented above should be valid also for realistic values of $U/t$ and $\alpha_R/h$.

There could in principle exist a transition to a magnetic state, such as the antiferromagnetic phase found for the 2D repulsive Hubbard model at half-filling in the weak-coupling limit.\cite{Hirsch1985} However, applying a Zeeman field splits the degenerate spin bands, and we therefore expect that no antiferromagnetic ordering can exist as long as $h > U$. Though the application of a Zeeman field could favor a ferromagnetic phase, other studies have indicated that ferromagnetic ordering does not appear in the weak-coupling limit of the 2D Hubbard model,\cite{Hirsch1985,Lin1987} a result we expect to hold also in the present case.

\section{Conclusion}\label{sec:conclusion}
We have investigated the role of a weak spin-orbit coupling on a spin-polarized weakly repulsive Hubbard system on a square lattice. Performing an analysis along the same lines as done by Lake \etal\cite{Lake2016} for the 2D electron gas, we found that the superconducting order was caused by the SOC-independent Kohn-Luttinger term in the interaction. The pairing symmetry was, however, indirectly determined by the SOC: the realized superconducting gap has the same chirality as the band. We also found that the momentum dependence of the superconducting gap could be tuned by rotating the magnetic field and changing the filling fraction. The filling fraction also determines which band has the highest $T_c$.

\begin{acknowledgments}
H.G.H. would like to thank F. N. Krohg for useful discussions.
This work was supported by the Research Council of Norway through Grant Number 250985 Fundamentals of Low-dissipative Topological Matter, and Center of Excellence Grant Number 262633, Center for Quantum Spintronics.
\end{acknowledgments}

\appendix*
\section{Expansion of susceptibility in square lattice harmonics}\label{sec:appendix}
\renewcommand{\theequation}{\Alph{section}\arabic{equation}}
To the order we are working, we can set $\alpha_R = 0$ when calculating the susceptibility, Eq.~(\ref{susceptibility}). In this case, the dispersion in Eq.~(\ref{eps_approx}), has the symmetries of the $C_{4v}$ group, and is invariant under spatial inverision, $\vk \rightarrow -\vk$, 4-fold rotations, $(k_x, k_y) \rightarrow (k_y, -k_x)$, mirror operations, $(k_x, k_y) \rightarrow (-k_x, k_y)$ etc. Since we in Eq.~(\ref{susceptibility}) sum over the 1BZ, it can be shown that the susceptibility has the same symmetries. The expansion of the susceptibility thus has to be invariant under the same operations, which greatly reduces the possible terms in the expansion. Since the susceptibility is even under inversions (only the SOC term breaks inversion symmetry, which is neglected here), the expansion must contain only even terms, which we write in a general form\cite{Otnes1997}
\begin{equation}
  \chi(\vq) = \sum_{m,n} a_{mn}\cos(mq_x + nq_y),\label{eq:A1}
\end{equation}
where $m$ and $n$ are integers, and the band index has been dropped for notational simplicity. From the requirement $\chi(q_x, q_y) = \chi(-q_x, q_y)$ we find $a_{mn} = a_{m,-n} = a_{-m,n}$, and similarily from $\chi(q_x, q_y) = \chi(q_y, -q_x)$ we find $a_{mn}=a_{-n,m} = a_{nm}$. Using these relations, we simplify the above equation:
\begin{equation}
    \chi(\vq)
    = a_{00} + \sum_{(m,n)>0}2a_{mn}[\cos(m q_x + n q_y) + \cos(m q_x - n q_y)].
\end{equation}
Seperating the terms according to if $m=n$ or not, we get
\begin{equation}
  \begin{aligned}
  \chi(\vq) = D_{00}G_{00} &+ \sum_{m>n>0} D_{mn}G_{mn}(\vq) \\
  &+ \sum_{m>0} [D_{0m} G_{0m}(\vq) + D_{mm} G_{mm}(\vq)],
\end{aligned}
\end{equation}
where we have redefined the expansion coefficients $a_{mn}$ and defined the orthonormal functions
\begin{subequations}
\begin{eqnarray}
    G_{00} &=& \frac{1}{2\pi},\\
    G_{0m}(\vq) &=& \frac{\cos m q_x + \cos m q_y}{2\pi},\\
    G_{mm}(\vq) &=& \frac{\cos m q_x \cos m q_y}{\pi},\\
    G_{mn}(\vq) &=& \frac{\cos m q_x \cos n q_y+ \cos n q_x \cos m q_y}{\sqrt{2}\pi}.
\end{eqnarray}
\end{subequations}
We now insert $\vq = \vk-\vk'$ and rewrite the above functions in terms of products of functions of $\vk$ or $\vk'$ separately,
\begin{eqnarray*}
    4\pi G_{0m}(\vk-\vk') &=& [(\sin mk_x + i\sin mk_y)(\sin mk_x' - i\sin mk_y') \\
    &+& \mathrm{h.c.}] + [\sin \rightarrow \cos],
\end{eqnarray*}
\begin{eqnarray*}
    \pi G_{mm}(\vk-\vk') &=& [(\cos mk_x \cos mk_y)(\cos mk_x' \cos mk_y')\\
    &+& (\cos mk_x \sin mk_y)(\cos mk_x' \sin mk_y')]\\
    &+& [\sin \leftrightarrow \cos],
\end{eqnarray*}
\begin{eqnarray*}
    \sqrt{2}\pi G_{mn}(\vk-\vk') &=& [(\cos mk_x \cos nk_y)(\cos mk_x' \cos nk_y')\\
    &+& (\cos mk_x \sin nk_y)(\cos mk_x' \sin nk_y')\\
    &+& (\sin mk_x \cos nk_y)(\sin mk_x' \cos nk_y')\\
    &+& (\sin mk_x \sin nk_y)(\sin mk_x' \sin nk_y')]\\
    &+& [n \leftrightarrow m].
\end{eqnarray*}
Since the SOC is weak, the interaction can be regarded to be between particles of equal spin to the order we are working. Hence, the interaction must be odd in $\vk$ and $\vk'$. In this way we can neglect most of the above terms, and are left with an expansion of the form
\begin{eqnarray}
  \chi(\vk-\vk') &=& \sum_{m} \chi^{0m}[g_{x+iy}(m\vk)g_{x-iy}(m\vk') + \mathrm{h.c.}] \nonumber\\
  &+&\sum_{m} \chi^{mm}[g_x(m \vk)g_x(m \vk') + g_y(m \vk)g_y(m \vk')]\nonumber\\
  &+& \sum_{m>n} \chi^{mn}\big[g_x(m k_x, n k_y)g_x(m k_x', n k_y') \nonumber\\
  &&\quad + g_y(m k_x, n k_y)g_y(m k_x', n k_y') + m\leftrightarrow n\big],
\end{eqnarray}
where $m,n>0$ and we have used the functions defined in Eq.~(\ref{expansion_functions}). The leading order terms included in Eq.~(\ref{chi_expansion}) correspond to the $\chi^{01}$, $\chi^{11}$ and $\chi^{21}$ terms in the above equation.

\end{document}